\documentclass[twocolumn]{aastex63}
\usepackage{graphicx}
\usepackage{mathtools}
\received{July 22, 2024}
\accepted{September 22, 2024}

\shorttitle{WASP-121A,\lowercase{b}}
\shortauthors{Sing et al.}
\graphicspath{{./}{figures/}}

\begin{document}

\title{An absolute mass, precise age, and hints of planetary winds for WASP-121 A and \lowercase{b} from a JWST NIRSpec phase curve}


\correspondingauthor{D. K. Sing}
\email{dsing@jhu.edu}

\author[0000-0001-6050-7645]{David K. Sing}
\affil{Department of Earth \& Planetary Sciences, Johns Hopkins University, Baltimore, MD, USA}
\affil{William H.\ Miller III Department of Physics \& Astronomy, Johns Hopkins University, 3400 N Charles St, Baltimore, MD, USA}

\author[0000-0001-5442-1300]{Thomas M. Evans-Soma}
\affiliation{School of Information and Physical Sciences, University of Newcastle, Callaghan, NSW, Australia}
\affiliation{Max Planck Institute for Astronomy, K\"{o}nigstuhl 17, D-69117 Heidelberg, Germany}

\author[0000-0003-4408-0463]{Zafar Rustamkulov}
\affil{Department of Earth \& Planetary Sciences, Johns Hopkins University, Baltimore, MD, USA}

\author[0000-0003-3667-8633]{Joshua D. Lothringer}
\affiliation{Space Telescope Science Institute, Baltimore, MD 21218, USA}

\author[0000-0001-6707-4563]{Nathan J. Mayne}
\affiliation{Department of Physics and Astronomy, Faculty of Environment Science and Economy, University of Exeter, EX4 4QL, UK}

\author[0000-0001-5761-6779]{Kevin C.\ Schlaufman}
\affil{William H.\ Miller III Department of Physics \& Astronomy, Johns Hopkins University, 3400 N Charles St, Baltimore, MD, USA}


\begin{abstract}

We have conducted a planetary radial velocity measurement of the ultra-hot Jupiter WASP-121b using JWST NIRSpec phase curve data. Our analysis reveals the Doppler shift of the planetary spectral lines across the full orbit, which shifts considerably across the detector ($\sim$ 10 pixels). Using cross-correlation techniques, we have determined an overall planetary velocity amplitude of $K_{\rm p}=215.7\pm1.1$ km/s, which is in good agreement with the expected value. We have also calculated the dynamical mass for both components of the system by treating it as an eclipsing double-line spectroscopic binary, with WASP-121A having a mass of M$_{\star}$=1.330 $\pm$ 0.019 M$_{\odot}$, while WASP-121b has a mass of M$_{\rm p}$= 1.170 $\pm$ 0.043 M$_{\rm Jup}$. These dynamical measurements are $\sim3\times$ more precise than previous estimates and do not rely on any stellar modeling assumptions which have a $\sim$5\% systematic floor mass uncertainty. Additionally, we used stellar evolution modeling constrained with a stellar density and parallax measurement to determine a precise age for the system, found to be 1.11 $\pm$ 0.14 Gyr. 
Finally, we observed potential velocity differences between the two NIRSpec detectors, with NRS1 lower by 5.5$\pm$2.2 km/s. We suggest that differences can arise from day/night asymmetries in the thermal emission, which can lead to a sensitivity bias favoring the illuminated side of the planet, with planetary rotation and winds both acting to lower a measured $K_{\rm P}$. The planet's rotation can account for 1 km/s of the observed velocity difference, with 4.5$\pm$2.2 km/s potentially attributable to vertical differences in wind speeds.

\end{abstract}




\section{Introduction} \label{sec:intro}

For transiting exoplanets, estimates of the stellar mass have predominantly relied on stellar evolution models constrained by measurements, including parallax, stellar radial velocity, and stellar density \citep{Hartman2019AJ....157...55H, Hellier2019MNRAS.490.1479H, Yee2023ApJS..265....1Y}. Alternatively, a planet's radial velocity signal can also be used and high-resolution ground-based infrared spectrographs have been used to detect exoplanet molecular features over the last decade, beginning with \cite{Snellen2010}, who used the CO lines to constrain the absolute stellar mass as the planetary Doppler shift was measured. As ground-based spectrographs have large systematics from sources such as telluric contamination and are limited in the uninterrupted duration they can observe, it is challenging to capture a complete phase curve of an exoplanet from the ground (see the review by \citealt{Birkby2018haex.bookE..16B} for additional details). Most such observations to date have concentrated around the transit or eclipse phase to measure the planet's spectrum, which is not an optimal phase to measure the planet's radial-velocity semi-amplitude and thus is not ideal for precision stellar mass measurements. However, some measurements do cover larger phase ranges (typically between phases 0.3 to 0.7) such as \cite{vanSluijs2023MNRAS.522.2145V} who covered WASP-33 b or \cite{Ridden2023AJ....165..211R} who measured KELT-9 b. These observations typically cover phase ranges between $\sim$ 0.3 to 0.75 which results in precise ($\sim$0.5\%) planetary semi-amplitudes.

JWST has enabled detailed spectral measurements of transiting exoplanets using transmission, emission, and phase curve spectroscopy (e.g., \citealt{Rustamkulov2022ApJ...928L...7R,Bean2023Natur.618...43B, August2023ApJ...953L..24A,Mikal-Evans2023ApJ...943L..17M,Bell2024NatAs...8..879B}). The infrared observatory has proven to be an extremely stable platform \citep{Rigby2023PASP..135d8001R, Espinoza2023PASP..135a8002E} with achievable photometric precisions on the order of 10s of parts-per-million or better (e.g., \citealt{Lustig-Yaeger2023arXiv230104191L,Coulombe2023Natur.620..292C}). The NIRSpec G395H grating, in particular, has shown to be extremely sensitive to the strong molecular absorption from CO$_2$ \citep{ERS2023Natur.614..649J, Rustamkulov2023Natur.614..659R} and CO \citep{Alderson2023Natur.614..664A}. As seen in the direct imaging planet VHS~1256-1257~b \citep{Miles2023ApJ...946L...6M} and the transiting planet WASP-39b \citep{Grant2023ApJ...949L..15G}, the fundamental CO bandheads between 4.4 and 5 $\mu$m are resolvable with G395H. Cross-correlation techniques have also been shown to be capable of detecting CO with NIRSpec/G395H \citep{Esparza-Borges2023ApJ...955L..19E}. With CO resolvable in long stare complete phase curve observations, JWST should be able to enable precision mass measurements if the planet's Doppler shift can be detected. 
Moreover, as demonstrated by Rustamkulov et al. (2023b), the precise stellar densities derived from the high signal-to-noise JWST transit light curves enable very tight constraints on stellar evolution models. 

WASP-121b is an ultra-hot Jupiter that is particularly favorable for atmospheric measurements \citep{Delrez2016MNRAS.458.4025D}. Various features have been detected in its atmosphere, including H$_2$O and Fe, through HST transmission, emission, and phase-curve measurements \citep{evans2016,evans2017,evans2018, Mikal-Evans2019MNRAS.488.2222M, Mikal-Evans2023ApJ...943L..17M, sing2019}. The planet has also been observed in optical transmission high-resolution spectra, with a large number of atomic species detected \citep{Gibson2020MNRAS.493.2215G, Bourrier2020A&A...635A.205B, Ben2020ApJ...897L...5B, Cabot2020MNRAS.494..363C, Merritt2021MNRAS.506.3853M, Borsa2021A&A...645A..24B, Azevedo2022A&A...666L..10A, Hoeijmakers2022arXiv221012847H, Wardenier2024PASP..136h4403W,Maguire2023MNRAS.519.1030M, Seidel2023arXiv230309376S}. Notably, the planet velocities measured from ESPRESSO transit measurements range from 197 to 213 km/s \citep{Maguire2023MNRAS.519.1030M}, which is significantly lower than the expected orbital velocity of the planet. Such a discrepancy could be a signature of winds or atmospheric escape. In addition, \cite{Wardenier2024PASP..136h4403W} found phase-dependent Doppler shifts in CO and H$_2$O around transit, due to a combination of planetary rotation and the spatial distribution of the molecular species.

In this work, we present new planetary radial velocity measurements of WASP-121b using the JWST/NIRSpec G395H instrument which has a spectral resolution near R$\sim$3000. 
This resolution is sufficient to observe significant wavelength shifts across the detector for short-period planets, given the large orbital velocities near $\sim$200 km/s compared to the NRS1 and NRS2 detector's average resolutions of 67 and 44 km/s per pixel, respectively.
We note for JWST the NIRSpec high resolution gratings will be needed for studies of this nature. For instance, with resolutions near 700 for NIRISS SOSS, the planet would only move across a couple of pixels peak-to-peak during the orbit, making it difficult to remove the stellar contribution cleanly.
We describe the data reduction in Section \ref{sec1}, our analysis in Section \ref{sec:analysis}, present our methods in Section \ref{sec:methods}, give results in Section \ref{sec:results}, and conclude in Section \ref{sec:concl}.


\section{Data Reduction}\label{sec1}
\begin{figure}
\begin{centering}
\includegraphics[width=0.5\textwidth]{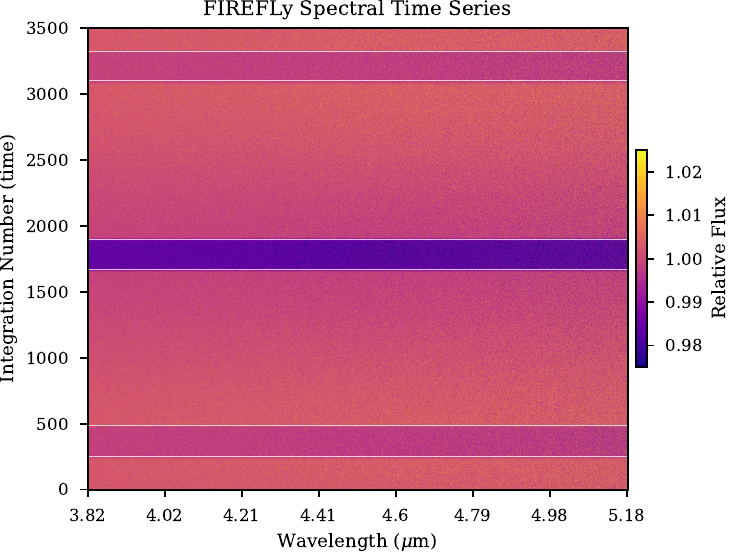}
\par\end{centering}
\vspace{-0.3cm}
	\caption{NIRSpec G395H NRS2 Spectrophotmetry. Plotted is the WASP-121 spectra vs detector integration number, which is proportional to time, with the color bar corresponding to the normalized flux. The transit can clearly be seen in the middle as well as the two eclipses at the top and bottom of the plot. The begining and end of the transits and eclipses are marked with white horizontal lines.} \label{fig:datacube}
\end{figure}

We use the phase curve measurements of WASP-121b taken with the JWST NIRSpec G395H instrument as part of GO-1729 (P.I. Mikal-Evans, co-P.I. Kataria). The data continuously covers the entire phase curve of the planet lasting 1.57 days, beginning just before secondary eclipse and ending shortly after a second secondary eclipse (see Fig. \ref{fig:datacube}). The NIRSPec/G395H grating covers wavelengths from $\lambda$ = 2.70 to 5.15$\mu$m at a resolution of R$\sim$3000. 

This dataset and the reductions used here have been described in \cite{Mikal-Evans2023ApJ...943L..17M}. In short, we use the \texttt{FIREFLy} suite to reduce the JWST data starting from the uncalibrated data \citep{Rustamkulov2022ApJ...928L...7R, Rustamkulov2023Natur.614..659R} optimizing the JWST Calibration pipeline for time series observations. The customized routines include removal of 1/f noise at the group and integration level, bad pixel removal, cosmic ray cleaning spatially, and an optimized extraction of the stellar spectrum.  

From the extracted Time Series Spectra (TSS), we removed stellar flux and planetary continuum flux from the data using the following procedure to isolate the planetary emission/absorption lines.  For each integration we:\\
\indent $\bullet$ Removed the white light phase curve contribution from each TSS, dividing the spectrophotometric timeseries at each wavelength by the wavelength integrated white light curve.\\
\indent $\bullet$ Divided-out the median spectra of the whole time series from each TSS. This procedure removes the stellar flux contribution and a portion of the planetary continuum, while preserving the Doppler-shifting planetary spectral lines. We note the Doppler shift of the stellar lines is well below the instrument resolution so the stellar lines are assumed to be stable throughout the phase curve observation.\\
\indent $\bullet$ Removed the phase-dependent planet continuum emission by dividing each TSS by a running median filter in wavelength chosen to be 101 pixels wide. This procedure removes any remaining phase-dependent planetary continuum flux while preserving individual spectral lines of the planet.

The residual spectrophotmetry can be seen in Fig. \ref{fig:datacubecleaned}, which we use in our cross-correlation analysis.
\section{Analysis} \label{sec:analysis}

\begin{figure}
\begin{centering}
\includegraphics[width=0.5\textwidth]{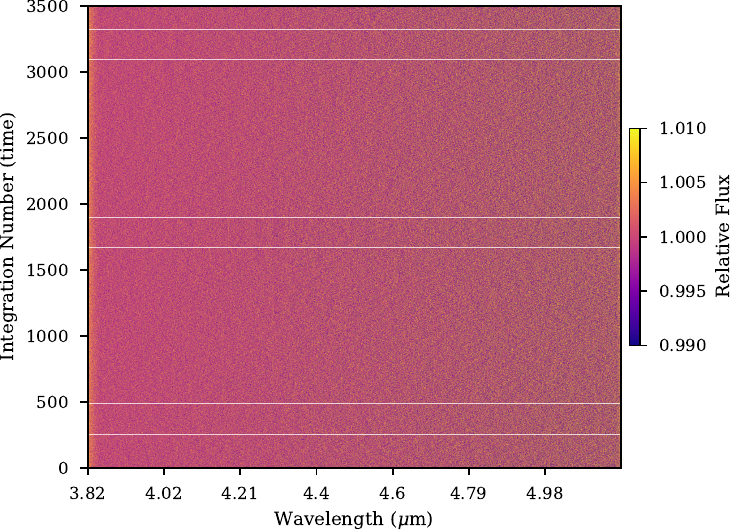}
\par\end{centering}
\vspace{-0.3cm}
	\caption{The same as Fig. 1 but plotting the residual spectrophotometry after removing the median stellar spectra, white light phase curve, and phase-dependent planetary broadband emission spectrum.} \label{fig:datacubecleaned}
\end{figure}

The residual data cube (with dimensions of time, $\lambda$, and flux), has the stellar contribution of the star effectively removed along with the planet's continuum flux, while the planetary spectral lines have been largely unaffected (see Fig. \ref{fig:datacubecleaned}). We note that the high quality of the JWST data allowed for cross-correlation measurements at this stage without the need for further cleaning or removal of further systematic errors in the data. Such additional steps are typically needed for ground-based high-resolution spectroscopy to remove, for instance, telluric contamination, airmass trends, or detector drifts (e.g. \citealt{Birkby2017AJ....153..138B}). In addition, the JWST phase curve data is a single continuous observation, such that the full orbital velocity curve can be measured without night-to-night or visit-to-visit calibration differences.

The wavelength calibration for JWST NIRSpec has been found to meet the requirement of 1/8 of a resolution element   \citep{Boker2023PASP..135c8001B}.  However, the accuracies assume a well-centered point source and the BOTS mode uses a wide slit which can lead to wavelength calibration offsets if the target is not well centered. Using stellar H lines, we compared the stellar spectra to a phoenix model at the known system velocity and found a wavelength shift over the default pipeline solution of about 1 pixel. Thus, a measurement of the system velocity with this dataset will have a large 1-pixel systematic uncertainty which prohibits a precise measurement. However, the relative planet velocities are preserved as the target remained well placed in the slit (1/500 of a pixel, $\sim$0.1 km/s) during the entire 37.8 hr phase curve observation \citep{Mikal-Evans2023ApJ...943L..17M}.
\subsection{Planet Velocity Measurement}
Assuming a circular orbit, the expected measured planet velocity, $v_p$, can be calculated from the stellar radius and transit-derived parameters using
\begin{equation}
v_p=\frac{2\pi(a/R_{\star})R_{\star}}{P}{\rm sin}(i),
\label{eq1}
\end{equation}
where $a/R_{\star}$ is the semi-major axis to stellar radii, $P$ is the period and $i$ the inclination. Using the values found in \cite{Bourrier2020A&A...635A.205B}, the expected planet velocity is calculated to be 220$\pm$4 km/s with the majority of the uncertainty due to the stellar radius. The G395H resolution varies from $R$=2000 to 3500 between 3 and 5 $\mu$m corresponding to velocity resolutions of 150 to 85 km/s or 68 to 38 km/s per pixel respectively. Thus, peak-to-trough the planetary signal is expected to shift across 6.5 pixels at 3 $\mu$m and 11.6 pixels at 5 $\mu$m. With a  shift $>$1 pixel, the planetary Doppler shift is expected to be readily detectable, especially when co-adding the signal of many spectral lines across the detector. 

To facilitate velocity measurements, using a cubic spline interpolation we re-sampled the G395H residual data cube spectra on a uniform log($\lambda$) wavelength scale and super-sampled each pixel by a factor of 10. With the log re-sampling, the spectra have a corresponding constant resolution in velocity for each detector of 6.83 km/s per pixel for NRS1 and 4.45 km/s per pixel for NRS2.

\subsubsection{Cross-correlation with forward models}
\begin{figure}
\begin{centering}
\includegraphics[width=0.5\textwidth]{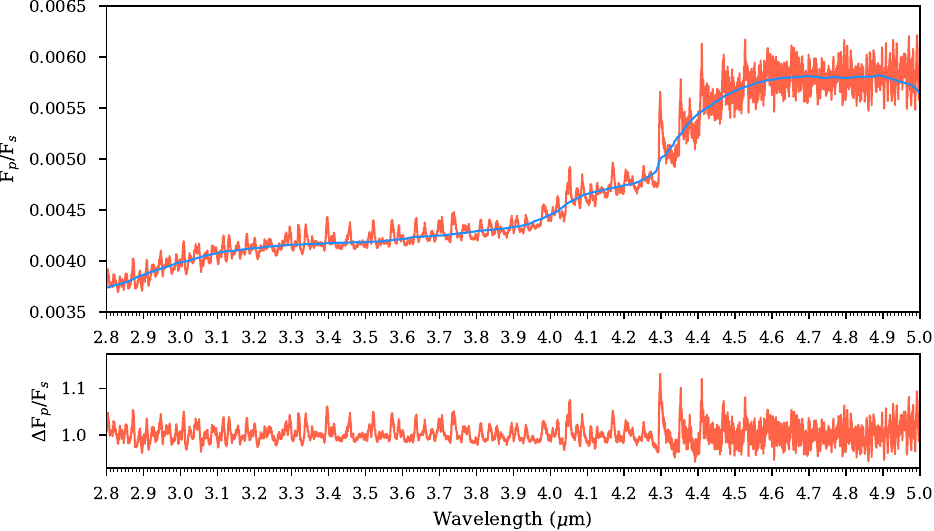}
\par\end{centering}
\vspace{-0.3cm}
	\caption{(Top) Phoenix forward model planetary spectrum corresponding to the dayside for WASP-121b. Emission lines from species such as CO are visible. (Bottom) the cross-correlation template Phoenix planetary spectrum generated by dividing by a median filter which removes the continuum emission.} \label{fig:phoenix}
\end{figure}

\begin{figure}[h!]
    \centering
        \centering
        \includegraphics[width=0.5\textwidth]{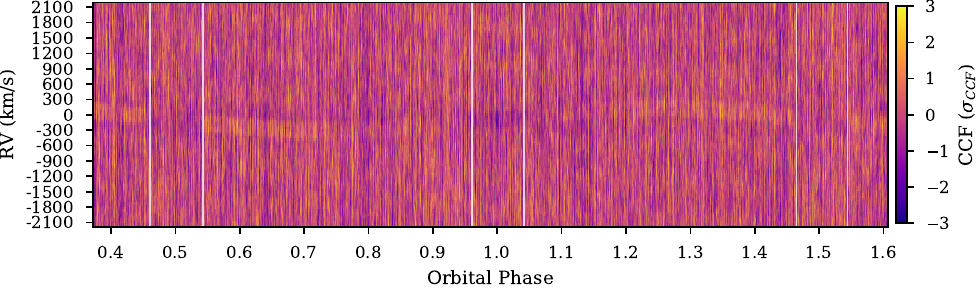}
        \label{fig:first}
    \vskip\baselineskip
    \vspace{-0.7cm}
        \centering
        \includegraphics[width=0.5\textwidth]{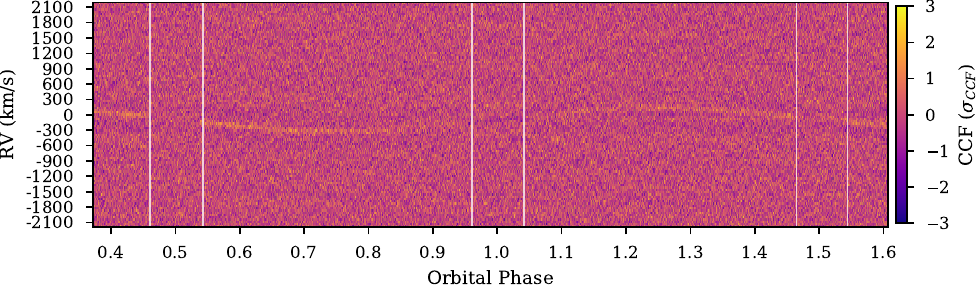}
        \label{fig:second}
        \vspace{-0.4cm}
    \caption{Cross correlation amplitude as a function of orbital phase and radial velocity lag using Phoenix  model as a correlation template. The CCF amplitude has been normalized at each phase by its standard deviation. (Top) Correlation using a Phoenix grid model as a correlation template. (Bottom) using a PHOENIX model template derived from a retrieval on the planetary dayside emission spectra. Vertical white lines mark transit and eclipse 1st and 4th contact phases.}
    \label{fig:combined}
\end{figure}

We cross-correlated the re-sampled residual spectral data with a PHOENIX model \citep{Lothringer2018ApJ...866...27L} representative of the dayside spectrum (see Fig. \ref{fig:phoenix}). The first model was an 'out-of-the-box' forward model generated from a grid, while the second was the best-fit emission spectra from a retrieval fit to the G395H dayside emission spectra (Mikal-Evans et al., in prep, priv. comm). Cross-correlation templates from the PHOENIX spectra were generated by removing the continuum with a median filter in wavelength as done for the data. The model template spectra has spectral emission features from CO and H$_2$O, which are expected based on previous measurements (e.g. \citealt{Mikal-Evans2019MNRAS.488.2222M}). We cross-correlated the model template with all 3504 spectra in the time series data.  The time dependent correlation results for both the 'out-of-the-box' and retrieval model spectra can be seen in Fig. \ref{fig:combined} showing the correlation as a function of radial velocity lag and orbital phase. A correlation is seen in the data detected at the phase and 220 km/s amplitude expected by the planetary velocity. For the 'out-of-the-box' template spectra, a correlation can be seen, but the correlation peak itself is lower by a factor of 2 compared to the retrieval model, and the width of the cross-correlation function was found 3 to 4x wider and notably flat-topped, prohibiting a precise velocity measurement. 
The dramatic difference between the model spectra highlights the sensitivity of cross-correlation techniques to the exact atmospheric model used and the difficulty of optimizing a signal if the atmospheric structure and composition is notably different than can be assumed from model grids.
We note the cross-correlation function can be seen to flip to negative correlation values when correlating against the transit signal, as a transit spectra has absorption lines which are darker in-transit compared to the continuum while inverted emission lines during secondary eclipse are brighter.

\subsubsection{Self-derived cross-correlation}
Rather than use a planetary atmospheric model for the template spectra to correlate the data against, we also derived the template from the data itself as is commonly done when analysing the radial velocity signatures of binary stars (e.g. \citealt{Sing2004AJ....127.2936S}). Using the planetary spectral features themselves as a template should, in principle, be fully optimal to measure relative velocity shifts as it avoids model miss-matches including missing or incomplete cross-sections and the assumed atmospheric structure and composition.
However, with a self-derived template the absolute systemic velocity is not directly measured. 
In addition, residual stellar features and detector artefacts could be present in the empirical template which might be picked up in the cross correlation at a low levels near zero velocities.
\begin{figure}
\begin{centering}
\includegraphics[width=0.47\textwidth]{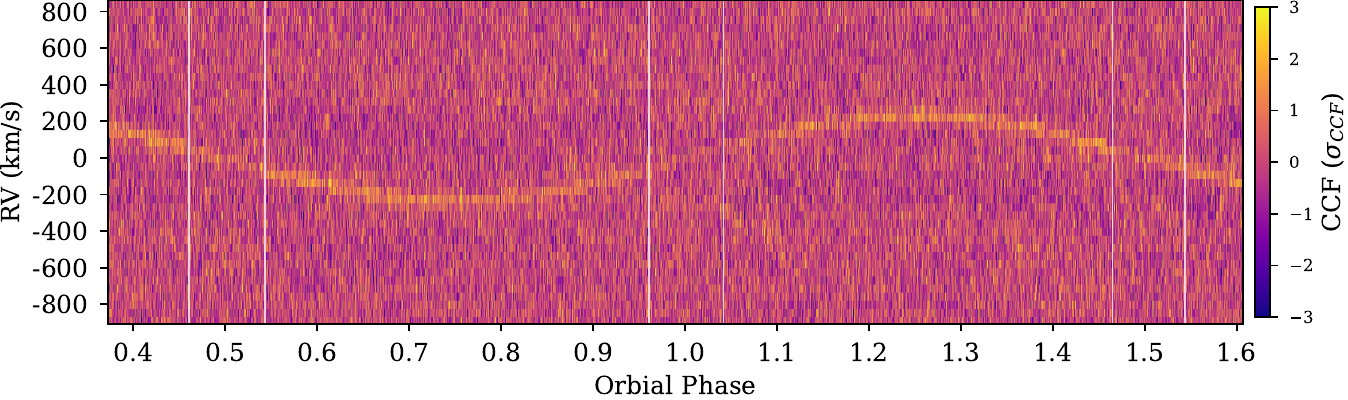}
\par\end{centering}
\vspace{-0.3cm}
	\caption{Same as Fig. \ref{fig:combined} but using a cross-correlation template derived from the data itself.} \label{fig:ccself}
\end{figure}

We first velocity shifted all the spectra to the expected planetary rest frame assuming $K_{\rm p}$=220 km/s, then computed the mean residual planetary spectrum of the entire residual time series. We then used the mean spectrum as a cross-correlation template, cross-correlating against the unshifted time series data. We normalized the cross-correlation signal strength by dividing the CCF function by its standard deviation. As seen in Fig. \ref{fig:ccself}, a planetary radial velocity signal is evident when plotting the cross-correlation signal as a function of orbital phase, with the correlation peaks much better resolved than when using the forward model. In order to time-resolve the planet signal at each phase at a 2-3 $\sigma$ level, we binned the spectra in time by a factor of 16 (corresponding 11.25 minute bins) and cross-correlated the binned spectra with our mean template. A Gaussian was then fit to strongest cross-correlation peak to determine the peak velocity shift and its uncertainty at each phase. At a few phases, the peak cross-correlation signal did not correspond to a velocity near the expected planet velocity. Given it occurred for only a small number of phases, we implemented a 15-$\sigma$ clip to remove those points. We additionally discarded the data during eclipse as the planetary emission is not observed during that time.

\begin{figure*}
\begin{centering}
\includegraphics[width=1.0\textwidth]{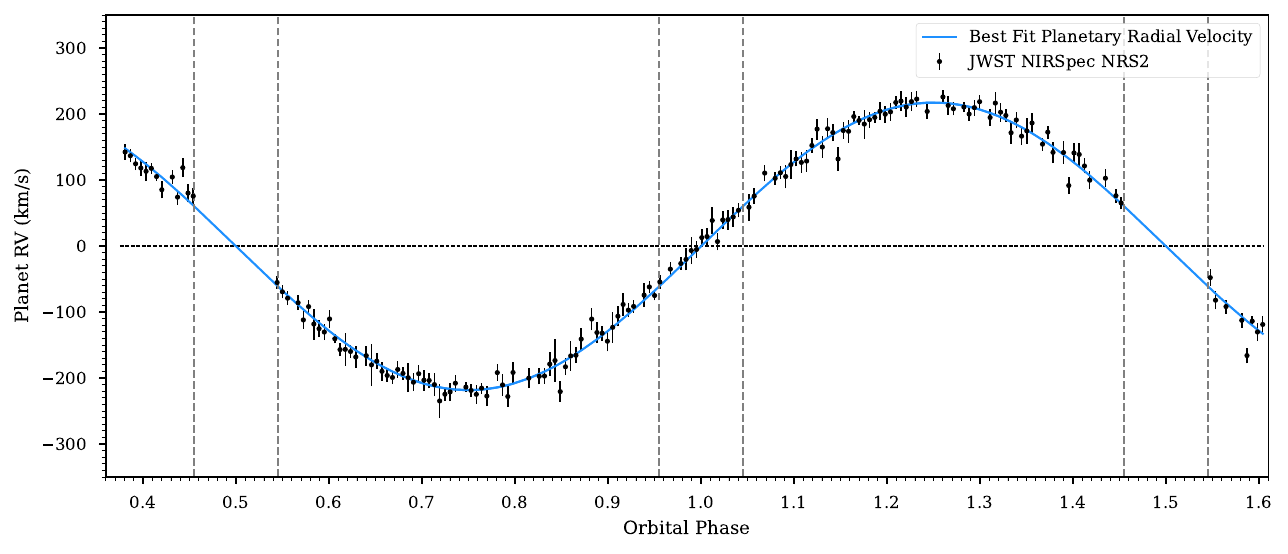}
\par\end{centering}
\vspace{-0.3cm}
	\caption{The planetary radial velocity signal derived using NIRSpec/NRS2 with 1-$\sigma$ uncertainties (black data points). A best-fit planetary radial velocity signal is also shown (blue) as well as the eclipse and transit first and forth contact phases (grey dashed lines).}
\label{fig:rvnrs2}
\end{figure*}

\begin{figure}
\begin{centering}
\includegraphics[width=0.45\textwidth]{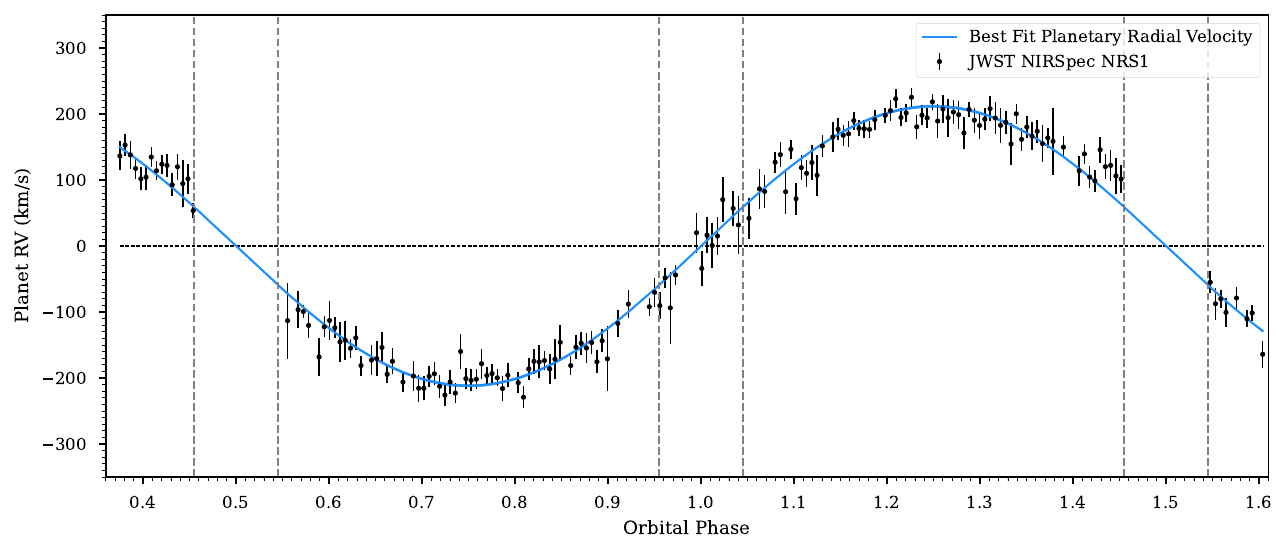}
\par\end{centering}
\vspace{-0.3cm}
	\caption{Same as Fig. \ref{fig:rvnrs2} but for NIRSPec/NRS1.} \label{fig:rvnrs1}
\end{figure}

For both NRS1 and NRS2, we fit the radial velocity (RV) curves with a sinusoid, fixing the period to that from \cite{Bourrier2020A&A...635A.205B} and fitting for the amplitude $K_{\rm p}$ and the absolute velocity (see Figs. \ref{fig:rvnrs2} and \ref{fig:rvnrs1}). Additionally fitting for a phase shift did not improve the fit, so no offset was assumed. To search for non-sinusoidal components, we additionally tried fitting the RV data with a high-order polynomial chosen to be a Taylor-expansion of a sinusoid, but did not find statistically significant results as measured by the Bayesian Information Criteria. For NRS1, we measure a velocity of $K_{\rm p}$ = $212.7\pm1.8$ km/s while $K_{\rm p}$ = $218.2\pm1.3$ km/s for NRS2. Although NRS2 has less total flux than NRS1, we find slightly higher precisions for NRS2 likely due to larger planetary emission along with correlating against stronger more favorable planetary features such as the comb-like CO lines. We report the weighted-mean value of $K_{\rm p}$ in Table \ref{tab:firefly_sys_params}. The radial velocity data is available on Zenodo (\url{https://doi.org/10.5281/zenodo.11992282}).

\section{Methods} \label{sec:methods}
\subsection{Dynamical Mass Measurements}
With the radial velocity of the star and planet both measured along with the inclination, we derive the masses for both components using the equation from \cite{Torres2010A&ARv..18...67T},
\begin{equation}
\begin{multlined}
M_{(\star,p)} \textrm{sin}^3(i) = \\
1.036149\times10^{-7}(1-e^2)^{3/2}(K_\star +K_P)^2 
K_{(p,\star)} P,
\end{multlined}
\end{equation}
where $M_{(\star,p)}$ is the mass of the star or planet (depending on which is being calculated), $K_\star$ is the stellar radial velocity and $e$ the eccentricity. Using  $K_\star$ from \cite{Delrez2016MNRAS.458.4025D}, both $e$ and $P$ from \cite{Bourrier2020A&A...635A.205B}, and a $K_{\rm p}$ value of $215.7\pm1.1$ km/s derived from our JWST data we find a stellar mass of $M_{\star}$=1.330$\pm$0.019 M$_\odot$ and $M_{p}$=1.170$\pm$0.043 M$_{\rm J}$.
We note that these dynamical mass measurements are about as precise as the best precisions found in the literature for this mass range, but do not rely on theoretical isochrones (e.g. \citealt{Hartman2019AJ....157...55H}). In addition, the precisions are improved by a factor of 2 to 3 over the recent measurements of WASP-121A,b from \cite{Bourrier2020A&A...635A.205B}. We report these values in Table \ref{tab:firefly_sys_params}, and include updated related parameters.

\subsection{Stellar evolution modeling}

We derive the fundamental and photospheric stellar parameters of WASP-121
using the \texttt{isochrones} \citep{mor15} package to execute with
\texttt{MultiNest} \citep{fer08,fer09,fer19} a simultaneous Bayesian
fit of the MESA Isochrones \& Stellar Tracks (MIST) isochrone grid
\citep{pax11,pax13,pax18,pax19,jer23,dot16,cho16} to a curated collection
of data for the star.  We fit the MIST grid to (1) SkyMapper Southern
Survey DR4 $uvgri$ photometry including in quadrature zero-point
uncertainties (0.03,0.02,0.01,0.01,0.01) mag \citep{onk24}, Gaia DR2
$G$ photometry including in quadrature its zero-point uncertainty
\citep{gai16,gai18,are18,bus18,eva18,rie18}, Two-micron All-sky Survey
(2MASS) $JHK_{s}$ photometry including their zero-point uncertainties
\citep{skr06}, and Wide-field Infrared Survey Explorer CatWISE2020 $W1 W2 W3$
photometry including in quadrature zero-point uncertainties (0.032,0.037,0.051)
mag 
\citep{wri10,mai11,mar21}; (2) a zero point-corrected Gaia DR3
parallax \citep{gai21,fab21,lin21a,lin21b,row21,tor21} with the formal uncertainty increased by 30\% \citep{El-Badry2021MNRAS.506.2269E}; and (3) an
estimated reddening value based on a three-dimensional reddening map
\citep{lal22,ver22}.  
For the GAIA photometry, we increased the error bar of the reported G-band photometry by 0.01 magnitudes, which takes into account the observed epoch-to-epoch scatter of the GAIA photometry, which we postulate as due to stellar activity.
We use a log uniform age prior between 0.1 Gyr and
10 Gyr, a uniform extinction prior in the interval 0 mag $< A_{V} < 0.2$
mag, and a distance prior proportional to volume between the \citet{bai21}
geometric distance minus/plus five times its uncertainty.  
The modeling is also constrained by a precise scaled semi-major axis value ($a/R_{\star}$) derived from the JWST transit light curve (Gapp et al. 2024, priv. comm.) which provides tight constraints on the stellar density.

\begin{figure}
\begin{centering}
\includegraphics[width=0.45\textwidth]{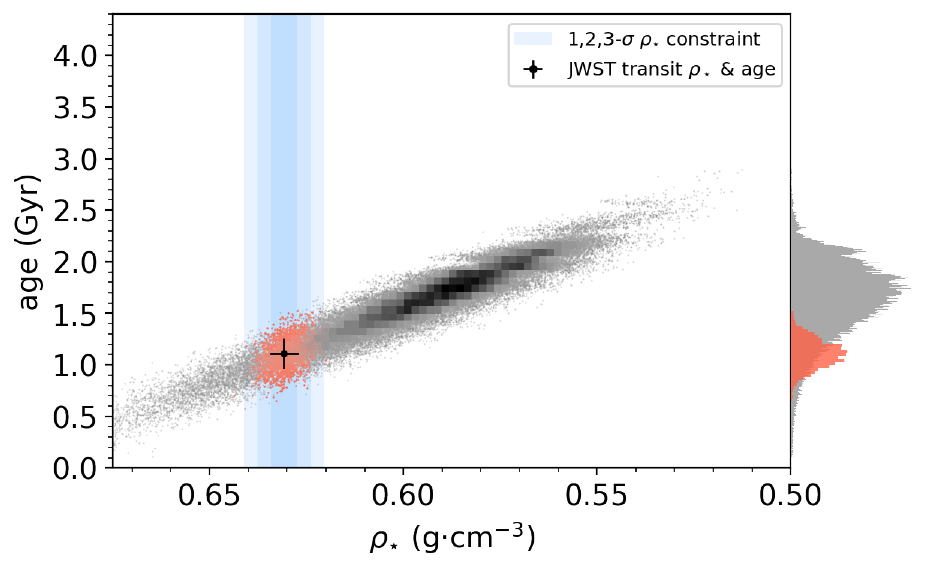}
\par\end{centering}
\vspace{-0.4cm}
	\caption{Isochrones posterior values of the age vs. stellar density considering only the broadband magnitudes, parallax and extinction (grey/black). An updated posterior with the added constraint of the JWST stellar density is also shown (red). The JWST measured stellar density and isochrone age is indicated with the black data point, with 1, 2, and 3-$\sigma$ density confidence intervals shown (blue).} \label{fig:corner}
\end{figure}

\begin{figure}
\begin{centering}
\includegraphics[width=0.45\textwidth]{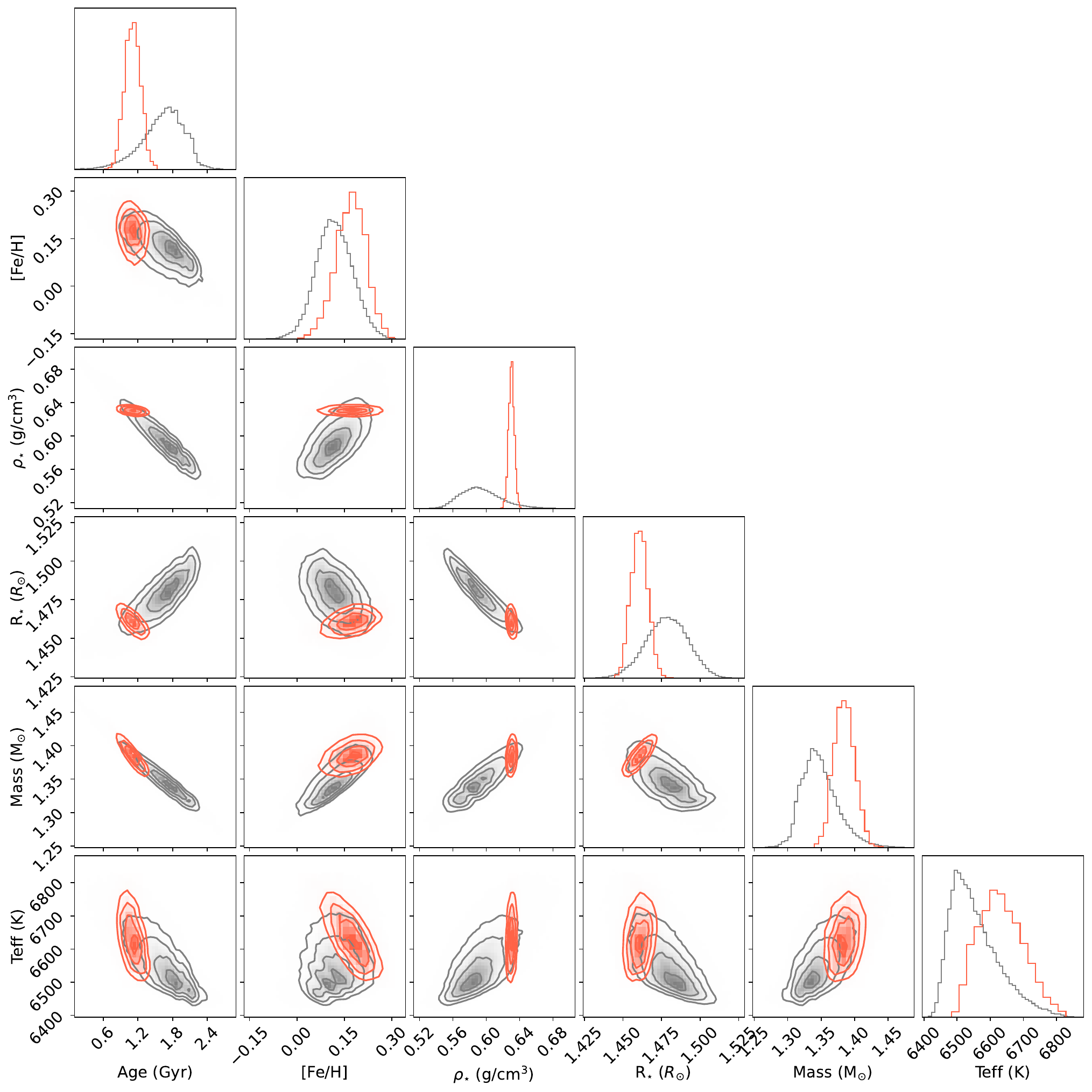}
\par\end{centering}
\vspace{-0.4cm}
	\caption{Isochrones posterior distribution of stellar parameters (grey/black) along with the posterior constrained with the JWST transit measured stellar density (red) with the values listed in Table \ref{tab:firefly_sys_params}.} \label{fig:fullcorner}
\end{figure}

\begin{figure}
\begin{centering}
\includegraphics[width=0.45\textwidth]{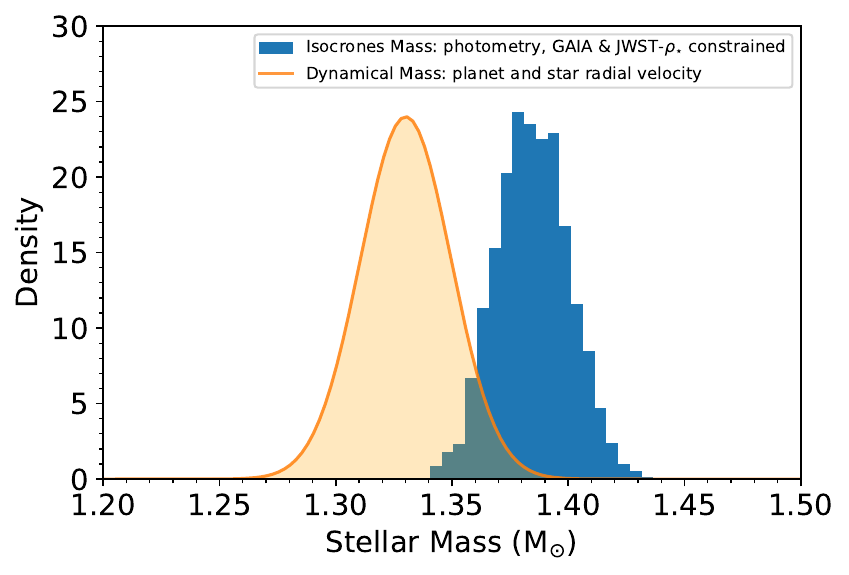}
\par\end{centering}
\vspace{-0.4cm}
	\caption{Dynamic radial velocity stellar mass measurement compared to the stellar modeling mass constrained by the broadband photometry, distance and stellar density.} \label{fig:masscomp}
\end{figure}

The results are shown in Figs. \ref{fig:corner} and \ref{fig:fullcorner} with the derived values in Table \ref{tab:firefly_sys_params}. We derive an age of 1.11$\pm$0.14 Gyr for the system. Our age is consistent with previous estimates \cite{Borsa2021A&A...645A..24B} though 3$\times$ more precise given the added constraint of the JWST transit measured stellar density. We also compare the isochrone-derived mass to the dynamical mass in Fig. \ref{fig:masscomp}, finding good agreement.

\begin{table}
    \centering
    \begin{tabular}{ccc} \hline
Parameter & Value \\ \hline
\hline
SkyMapper u & 11.7994$\pm$0.0230\\
SkyMapper v & 11.3884$\pm$0.0200\\
SkyMapper g & 10.5448$\pm$0.0146 \\
SkyMapper r & 10.3774$\pm$0.0149 \\
SkyMapper i & 10.3499$\pm$0.0151 \\
Gaia G DR2 & 10.3746$\pm$0.01\\ 
2MASS J & 9.625$\pm$0.021\\
2MASS H & 9.439$\pm$0.025 \\
2MASS Ks & 9.374$\pm$0.022 \\
WISE W1 & 9.365$\pm$0.033 \\
WISE W2 & 9.387$\pm$0.038 \\
WISE W3 & 9.383$\pm$0.062\\
Parallax (mas) & 3.8114$\pm$0.0135\\
A$_{\rm v}$ & 0.1118$\pm$0.0013\\
\hline

\end{tabular}
    \caption{WASP-121A isochrone inputs}
    \label{tab:stellar_params}
\end{table}

\section{Results} \label{sec:results}

\subsection{Tidally Mediated Orbital Evolution}
Unlike most hot Jupiter systems, the orbital period of the WASP-121
system $P = 1.27492504 \pm 0.000000145$ days is longer than the
rotation period of the star WASP-121 $P_{\text{rot}} \approx
1.13$ days \citep{Delrez2016MNRAS.458.4025D}.  As a consequence,
if WASP-121's stellar obliquity $\psi$ was identically zero, then
angular momentum would move from stellar rotation to the orbit of the
star--planet system thereby increasing its orbital period and semimajor
axis.  WASP-121 has stellar obliquity $\psi = 88.1^{\circ}$ though
\citep{Bourrier2020A&A...635A.205B}, necessitating a more comprehensive
treatment of the system's tidal evolution.  We therefore use the
tidal evolution model outlined in \citet{Leconte2010A&A...516A..64L}
to predict the instantaneous change in WASP-121's orbital period due
to star--planet tidal interactions.  Using (1) $a$, $M_{\star}$,
$R_{\star}$, $M_{\text{p}}$, and $R_{\text{p}}$ reported in Table
\ref{tab:firefly_sys_params}; (2) $P_{\text{rot}} = 1.13$ days
from \citet{Delrez2016MNRAS.458.4025D}; (3) $\psi = 88.1^{\circ}$
from \citet{Bourrier2020A&A...635A.205B}; (4) a stellar moment of
inertia $I/M_{\star} R_{\star}^2 = 0.056$ based on models presented in
\citet{Amard2019A&A...631A..77A} for the evolution of rotating stars; 
and (5) assuming $e = 0$ and modified planetary tidal quality factor
$Q_{\text{p}}' = 10^{5}$ as appropriate for a giant planet, we can
predict $\dot{P}$ for the WASP-121 system as a function of modified
stellar tidal quality factor $Q_{\star}'$.

We find $\dot{P} = -8.5 \times 10^{-6} \left(10^{8}/Q_{\star}'\right)$
s yr$^{-1}$.  Even if WASP-121 is highly dissipative like WASP-12 with
$Q_{\star}' \sim 10^{5}$, then $\dot{P} = -8.5$ ms yr$^{-1}$ a factor
of about three less than the $\dot{P}$ observed in the WASP-12 system
\citep[e.g.,][]{Yee2020ApJ...888L...5Y}.  While the value of $Q_{\star}'$
for WASP-121 is uncertain, \citet{Weinberg2024ApJ...960...50W} 
calculated $Q_{\star}'$ as a function of stellar mass, stellar age,
system orbital period, and planet mass.  For systems like WASP-121,
they predict $Q_{\star}' \sim 10^{8}$ that implies a vanishingly small
and certainly undetectable $\dot{P} = -8.5$ $\mu$s yr$^{-1}$.

\subsection{Planetary Wind Constraints}
We find a tentative 
$\Delta K_{NRS2-NRS1}$=$5.5\pm2.2$ km/s 
difference in the radial velocity (RV) measurement between the NRS1 and NRS2 detectors (2.5-$\sigma$ confidence). In particular, the measured velocity of NRS1 at 212.7$\pm$1.8 km/s is slower than the expected planetary orbital velocity of 217.8$\pm$1.0 km/s (calculated using Eq. \ref{eq1} with the updated values from Table \ref{tab:firefly_sys_params}).

\begin{figure}
\begin{centering}
\includegraphics[width=0.45\textwidth]{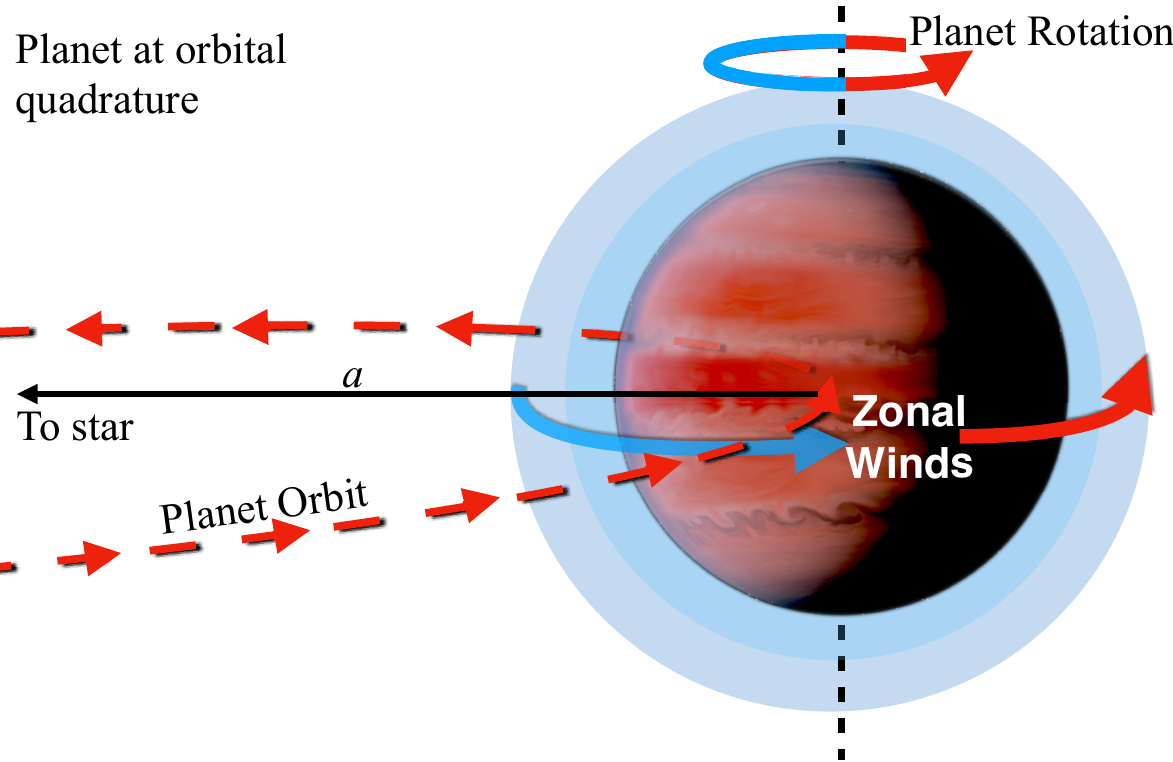}
\par\end{centering}
\vspace{-0.4cm}
	\caption{Illustration of the planet at orbital quadrature (phase 1.25) and the red/blue shifted velocity and day/night components.  At this phase, the planet is red-shifted from the orbital velocity,  the dayside emission of the planet is blue-shifted from winds and rotation while the fainter night-side is redshifted from winds and rotation. } \label{fig:rv}
\end{figure}

Unlike RV measurements of the star, the tidally-locked planet has large day/night temperature differences which can affect the measurements. In the absence of such differences, the velocity shifts imparted from the rotation of the planet and global equatorial winds will largely cancel out. With strong day/night differences, the planet can induce RV signals beyond the orbital velocity itself. For instance, in a limiting case where the night-side flux is negligible and the planetary emission emanates entirely from  the hot day-side, the tidally-locked rotation and equatorial winds will both act to reduce the measured radial velocity (see Fig. \ref{fig:rv}).

We performed an analytical estimate to better quantify the contributions from winds and the planet rotation on the NRS1/NRS2 velocity difference, concentrating first on relative velocity differences. We estimated the values at quadrature, given that phase has the maximum velocity signature and the planet can easily be divided geometrically into equal day and night-side components. With a 5.5 km/s difference observed between the detectors, we estimated what fraction of that difference could be directly attributable to the planet's rotation. At the equator, we calculate a planetary rotational velocity of $K_{rot}=7.6$ km/s derived from the planet's radius and orbital period (assuming the planet is tidally locked). 
As the bulk of planetary flux will be emitted from the equatorial region which is expected to be hottest \citealt{Mayne2014A&A...561A...1M, Parmentier2018A&A...617A.110P, Lee2022MNRAS.517..240L}, we estimated the rotational velocity components along the equator at the average angle emitted from the planet ($\mu=\textrm{cos}(\theta$)=1/2) or $\theta=60^\circ$.  
We flux-weighted the day and night-side contributions, $F_d$ and $F_n=1-F_d$ respectively, from the bulk planetary rotation taking the projected $\textrm{sin}(\theta)$ component contributing to the measured Doppler shifts giving:
\begin{equation}
\Delta K_{rot} = (1-F_d) K_{rot} \textrm{sin}(\theta)+ F_d K_{rot}\textrm{sin}(\theta).
\end{equation}
For NRS1, $F_d$=96\% as measured by \cite{Mikal-Evans2023ApJ...943L..17M} giving an estimated RV shift of 6.0 km/s while $F_d$=88\% for NRS2 giving an estimated RV shift of 5.0 km/s. Thus, the RV difference between the two detectors which can be attributable to the planetary rotation should be 1 km/s. This implies that the wind speed differences between the pressures probed between the two detectors is 4.5$\pm$2.2 km/s.
These estimates are in-line with current predictions  \citep{Lee2022MNRAS.517..240L} which find speeds up to 8 km/s and large vertical differences in windspeeds.

The NRS1 and NRS2 detectors are expected to probe different pressure levels, as the chemical and thermal differences from equator to pole and from the day to night side mean that at constant wavelength the pressure and height in the atmosphere one is detecting at a given optical depth varies.
NRS1 is expected to be sensitive to H$_2$O in particular while NRS2 at longer wavelengths is sensitive to strong CO and CO$_2$ lines (e.g. \citealt{Rustamkulov2022ApJ...928L...7R}). These pressure and longitude differences can result in differing wind profiles.
Comparing the absolute velocities measured, the velocity from NRS2 of 218.2$\pm$1.0 km/s matches the expected velocity of 217.8$\pm$1.0 km/s calculated using the JWST transit-derived $a/R_{\star}$ and $R_{\star}$ constrained using GAIA. With a $\sim$5 km/s shift expected from the planet's rotation not observed, this implies a counter-rotating wind component is largely able to cancel out this expected shift. The wind-patterns from GCM modeling predict such counter-rotating winds contribute at high latitudes \citep{Lee2022MNRAS.517..240L}. 
The lower RV value inferred from the NRS1 observations, relative to NRS2, suggests either that the prograde wind (i.e. flowing in the direction of the planet's rotation) contributes more substantially to the signal, or that the windspeeds at high latitudes are slower.
De-projected, the non-rotational 4.5 km/s RV difference suggests zonal windspeeds on the order of 4.5/$\textrm{sin}(\theta)$= 5.2 km/s.

A more sophisticated 3D treatment which takes into account the full viewing geometry, temperature structure, and atmospheric composition will be need to interpret the winds using this technique.  However, this estimate indicates that the winds can be probed with JWST  phase curve emission data given the sensitivity to the Doppler shifts from the global planetary emission  lines. Further, isolating and comparing specific spectral lines from the emission spectra (e.g. CO vs H$_2$O) can lead to further insights, however, such studies are beyond the scope of this work. 

These measurements nicely complement the existing high-resolution optical transit and phase curve measurements (e.g. \citealt{Maguire2023MNRAS.519.1030M,Hoeijmakers2022arXiv221012847H}), which probe higher altitudes in atomic transitions. In particular, several species including Fe have been detected at NUV and optical wavelengths and at altitudes beyond the Roche limit \citep{sing2019,Maguire2023MNRAS.519.1030M}. The phase curve measurements here are probing molecular emission near the mbar to bar region, which can give insight into the large equatorial jets expected on these planets. As the GCM wind speeds are highly dependent on the modeled diffusion \citep{Cooper2005ApJ...629L..45C, Mayne2014A&A...561A...1M}, comparing the theoretical wind speed values to the observations should help calibrate these types of models.  The large wind speeds estimated here are an indication the equatorial jets can be driven to high velocities in ultra-hot Jupiters.
We note that the significance (2.4-$\sigma$) is similar to that of the first windspeed measurement from CO on HD~209458b \citep{Snellen2008}. Given the potential differences between species, (e.g. \citealt{Wardenier2024PASP..136h4403W}), these estimates may be further refined by studies isolating specific molecular features and further emission/transmission spectral measurements which will improve the atmospheric constraints and therefore the correlation with model atmospheres.


\begin{table*}[t]
    \centering
    \begin{tabular}{ccccc} \hline
Parameter & Description & Units & Value & Reference(s)\\ \hline
\hline
\multicolumn{4}{l}{Measured planet parameters}\\
$K_{\rm p}$ &  RV semi-amplitude & km s$^{-1}$ & 215.7 $\pm$ 1.1 & this work \\
$P$ & orbital period & days & 1.27492504 $\pm$ 0.00000015 & Bourrier et al. (2020)\\
$a/R_\star$ & scaled semi-major axis &  & 3.7844 $\pm$ 0.0069 & Gapp et al. (2024)\\
$R_\mathrm{p} / R_\star$ & planet-star radius ratio & & 0.122551 $\pm$ 0.000063 & Mikal-Evans et al. (2023) \\
$(R_\mathrm{p} / R_\star)^2$ & transit depth & & 0.015018 $\pm$ 0.000015 & Mikal-Evans et al. (2023) \\

\multicolumn{4}{l}{Measured stellar parameters}\\
$K_\star$ & RV semi-amplitude & m s$^{-1}$ & 181.1 $\pm$ 6.4 & \cite{Delrez2016MNRAS.458.4025D} \\
$\rho_\star$ & stellar density & g cm$^{-3}$ &   0.6308 $\pm$ 0.0034  & this work \\
$d$ & distance & pc &  263.18 $\pm$ 0.72  & \cite{GAIA_dr3}\\

\multicolumn{1}{l}{Inferred planet parameters} \\
$T_\mathrm{eq}$ & equilibrium temperature & K & 2409 $\pm$ 24 & this work \\
$a$            & semi-major axis        & au & 0.02571 $\pm$ 0.00010 & this work\\
$R_\mathrm{p}$ & planet radius & R$_\mathrm{J}$& 1.7420 $\pm$0.0060  & this work \\
$\rho_\mathrm{p}$ & planetary density & g cm${^{-3}}$ &  0.275$\pm$0.010 & this work \\

\multicolumn{4}{l}{Dynamically determined parameters} \\
$M_\mathrm{p}$ & planet mass & M$_\mathrm{J}$& 1.170 $\pm$ 0.043 & this work \\
$M_\star$ & stellar mass & M$_\odot$ & 1.330 $\pm$ 0.019  & this work \\

\multicolumn{4}{l}{Inferred isochrones stellar parameters}\\
$M_\star$ & stellar mass & M$_\odot$ & 1.385 $\pm$ 0.016  & this work \\
$R_\star$ & stellar radius & R$_\odot$ & 1.461 $\pm$ 0.005 & this work \\
log($g$) & surface gravity & log10(cm s$^{-2}$) & 4.251 $\pm$ 0.003 & this work \\
$T_\mathrm{eff}$ & effective temperature & K & 6628 $\pm$ 66  & this work \\
$[$Fe/H] & metallicity & log([Fe/H]$_\odot$) & 0.17 $\pm$ 0.05  & this work \\
$\tau_{\mathrm{iso}}$ & isochronal age & $\times$10$^9$ yr & 1.11 $\pm$ 0.14 & this work \\
\hline

\end{tabular}
    \caption{System parameters for WASP-121 A,b.}
    \label{tab:firefly_sys_params}
\end{table*}

\section{Conclusion} \label{sec:concl}
We have presented a radial velocity measurement for the planet WASP-121b, which we detected during the entire orbital phase using JWST NIRSpec. Such measurements require the high-resolution gratings of NIRSpec, which are capable of detecting the orbital motion of close-in planets during phase curve observations.

With NIRSpec, we measured the radial velocity of the planet to 0.5\% precision, with $K_{\rm p}$=215.7$\pm$1.1 km/s. This is the first time an exoplanet's radial velocity has been continuously measured during an entire orbit. With both the star and planet semi-amplitude radial velocity detected, the absolute masses of both components were precisely determined. We found that WASP-121A has a mass of M$_{\star}$=1.330 $\pm$ 0.019 M$_{\odot}$, while WASP-121b has a mass of M$_{\rm p}$= 1.170 $\pm$ 0.043 M$_{\rm Jup}$. These masses are 2 to 3 times more precise than previous estimates and do not rely on stellar evolution models.

In the case of WASP-121b our radial velocities measure the mass of the star to 1.4\%, the JWST transit measures the stellar density to 0.5\%, and GAIA parallax and photometry measures the radius to 0.35\%. All of these precise independent stellar measurements are consistent at the 1-$\sigma$ level. 
As a result, the planetary mass and radius parameters are also precise and can be derive free of stellar modelling. 
Our 1.4\% stellar mass uncertainty improves upon the $\sim$5\% inherent uncertainty from estimating stellar masses using stellar models \citep{Tayar2022ApJ...927...31T}, with our data providing valuable calibration information.
In the case of WASP-121A, the good match to the stellar evolution models indicates the MESA models are accurate for this stellar type. This assessment agrees with results benchmarking the models to an open cluster \citep{Brandner2023MNRAS.518..662B}. 
There are a number of targets where these measurements can be made, as a number of planets have either comparable or higher emission spectral signals or higher amplitude planetary radial velocities (e.g. TOI-2109 b expected to be 291 km/s). 

Using stellar evolution models constrained with a precise JWST stellar density derived from the NIRSpec transit light curve, we found that the system is 1.11$\pm$0.14 Gyr old. This age makes the star only in the first quarter of its main sequence lifetime.
Our JWST planetary RV measurements on a K=9.3 magnitude star were able to reach median precisions of 11 km/s in a 11.25 min observation and 1 km/s for the whole 1.57 day phase curve. Future observers should consider phase curve measurements with the high resolution NIRSpec, as precision absolute masses can be obtained for both the star and planet, independent of stellar evolution models and constraining the zonal winds appears feasible. The large shifts in the planetary signal ($\sim$10-pixels) also highlights that the planetary RV will have to be taken into account when retrieving the atmospheric properties from the whole phase-curve signal. Otherwise, key molecular lines will not align in wavelength with the model.

Finally, we observed a potential wavelength dependence to the radial velocity amplitude, with the shorter wavelength NRS1 detector lower by 5.5$\pm$2.2 km/s. We estimated that the planet's rotation can account for only 1 km/s of the difference, with 4.5$\pm$2.2 km/s attributable to average zonal windspeeds tentatively estimated to be about 5.2 km/s.


\section*{ACKNOWLEDGEMENTS}
This work is based on observations made with the NASA/ESA/CSA James Webb Space Telescope. The data were obtained from the Mikulski Archive for Space Telescopes at the Space Telescope Science Institute, which is operated by the Association of Universities for Research in Astronomy, Inc., under NASA contract NAS 5-03127 for JWST. These observations are associated with program 1729. The data presented in this article were obtained from the Mikulski Archive for Space Telescopes (MAST) at the Space Telescope Science Institute. The specific observations analyzed can be accessed via \dataset[10.17909/6qnn-6j23]{https://doi.org/DOI}.
Support for JWST program GO-1729 was provided by NASA through a grant from the Space Telescope Science Institute, which is operated by the Association of Universities for Research in Astronomy, Inc., under NASA contract NAS 5-26555. 
J.K.B. was supported by a Science and Technology Facilities Council Ernest Rutherford Fellowship. 
N.J.M. acknowledges support from a UKRI Future Leaders Fellowship [Grant MR/T040866/1], a Science and Technology Facilities Funding Council Nucleus Award [Grant ST/T000082/1], and the Leverhulme Trust through a research project grant [RPG-2020-82].
N.M. was partly supported by a Science and Technology Facilities Council Consolidated Grant [ST/R000395/1], the Leverhulme Trust through a research project grant [RPG2020-82] and a UKRI Future Leaders Fellowship [grant number MR/T040866/1].
This work has made use of data from the European Space Agency (ESA) mission
{\it Gaia} (\url{https://www.cosmos.esa.int/gaia}), processed by the {\it Gaia}
Data Processing and Analysis Consortium (DPAC,
\url{https://www.cosmos.esa.int/web/gaia/dpac/consortium}). Funding for the DPAC
has been provided by national institutions, in particular the institutions
participating in the {\it Gaia} Multilateral Agreement.

\software{NumPy \citep{numpy}, SciPy \citep{scipy}, MatPlotLib \citep{matplotlib}, AstroPy \citep{astropy}}
\facilities{ JWST(NIRSpec) }
\section*{Data Availability} The radial velocity data products as shown in Figures \ref{fig:rvnrs2} and \ref{fig:rvnrs1} are available on Zenodo at  \url{https://doi.org/10.5281/zenodo.11992282}.
\bibliography{wasp121pcrv}{}
\bibliographystyle{aasjournal}

\end{document}